\documentclass{aa}  

\usepackage{graphicx}
\usepackage{txfonts}
\usepackage{color}

\def\gsim{\mathrel{\raise.5ex\hbox{$>$}\mkern-14mu
             \lower0.6ex\hbox{$\sim$}}}
\def\lsim{\mathrel{\raise.3ex\hbox{$<$}\mkern-14mu
             \lower0.6ex\hbox{$\sim$}}}

\begin{document}

	\title{The 3D pulsar magnetosphere with machine learning: first results} 
 \titlerunning{Pulsars with PINNs}
\authorrunning{Dimitropoulos, Chaniadakis, Contopoulos}
	
	\author{Ioannis Dimitropoulos\inst{1,2}, Evangelos Chaniadakis\inst{3}
		\and Ioannis Contopoulos
		\inst{2}
	}
	
	\institute{Department of Physics, University of Patras, Rio 26504, Greece\\
		\email{johndim888@gmail.com}
		\and
		Research Center for Astronomy and Applied Mathematics, Academy of Athens, Athens 11527, Greece\\
	    	 \email{icontop@academyofathens.gr}
		\and
		Department of Electrical Engineering, National Technical University of Athens, Athens 15772, Greece}
	
	
	
	\abstract
	{
\textcolor{black}{All PIC simulations of the pulsar magnetosphere over the past decade} show closed-line regions that end a significant distance inside the light cylinder, and manifest thick strongly dissipative separatrix surfaces instead of thin current sheets, with a tip that has a distinct pointed Y shape instead of a T shape. We need to understand the origin of these results which were not predicted \textcolor{black}{by our earlier numerical simulations} of the pulsar magnetosphere.
}
	{In order to gain new intuition on this problem, we set out to obtain the theoretical steady-state solution of the 3D ideal force-free magnetosphere with zero dissipation along the separatrix and equatorial current sheets. In order to achieve our goal, we needed to develop a novel numerical method. 
}
	{We solve two independent magnetospheric problems without current sheet discontinuities in the domains of open and closed field lines, and adjust the shape of their interface (the separatrix) to satisfy pressure balance between the two regions. The solution is obtained with meshless Physics Informed Neural Networks (PINNs).}
	{In this paper we present our first results for an inclined dipole rotator using the new methodology. We are able to zoom-in around the Y-point and inside the closed-line region, and we observe new interesting features. This is the first time the steady-state 3D problem is addressed directly, and not through a time-dependent simulation that eventually relaxes to a steady-state.}
	{We have trained a Neural Network that instantaneously yields the three components of the magnetic field and their spatial derivatives at any given point. {Our results demonstrate the potential of the new method to generate new solutions of the ideal pulsar magnetosphere}. 
}
	
    \maketitle
	%

\section{Gaps in our current understanding}

Right after the discovery of pulsars \citep{1968SciAm.219d..25H}, Goldreich and Julian formulated the mathematical description of their plasma-filled magnetospheres in the ideal force-free limit in steady-state and axisymmetry \citep{1969ApJ...157..869G}. It took us three decades to produce the first solution of this idealized problem, mainly because of the mathematical singularity along the light cylinder \citep{1999ApJ...511..351C}. Since then, this solution became a reference solution against which the results of numerical simulations are evaluated. As we will see below, the ideal force-free steady-state 3D problem is also well formulated mathematically, but still lacks a reference solution. This is mainly due to the presence of electric current sheets within the magnetosphere which strongly complicate all our standard numerical approaches\footnote{\textcolor{black}{In axisymmetry, the current sheet may be fixed to lie along the boundary of the computational domain in the equator as in \cite{1999ApJ...511..351C}, while in 3D the position of the current sheet must be determined by the simulation. If one decides to simulate the full axisymmetric magnetosphere as in \cite{Cerutti2015}, the same complication arises since the current sheet is free to oscillate around the equator.}}.


Recent advances in global particle-in-cell (PIC) simulations \citep[e.g.][]{2015ApJ...801L..19P,2022ApJ...939...42H,2023ApJ...943..105H,2023ApJ...958L...9B,2024A&A...690A.170S}  yield separatrix regions between open and closed field lines that exhibit a significant thickness beyond the simulation resolution (see Appendix), and a multi-layered internal structure. This is intriguing because, the same simulations are able to generate Harris-type equatorial current sheets with thickness at the limit of their numerical resolution (the skin-depth). Moreover, their closed-line regions terminate well inside the light cylinder, with regions of strong electromagnetic dissipation beyond their tips, something that was not expected by Goldreich and Julian. Interestingly, the shape of the tips resembles a pointed Y, thus deviating from the theoretically predicted T shape \citep{2003ApJ...598..446U}. Furthermore, the extent of the closed-line region and the sharpness of the Y appears to be linked to the efficiency of pair production in the simulation \citep[see comment in][]{2018ApJ...857...44K}. Finally, our recent work in \citet{2024Univ...10..178C} indicates that as the closed-line region approaches the light cylinder, the electric current in the 
 decreases, and in the limit where it touches the light cylinder, the equatorial current sheet vanishes entirely.

The origin of these results remains unclear. Some researchers suspect that they arise either due to insufficient resolution in current PIC simulations, or to inadequate titletimes to reach a true steady-state. 
On the other hand, the resolution of current PIC simulations falls significantly short in modeling the microphysics of the equatorial current sheet. To generate pulsar light curves and spectra for comparison with observations, simulation results are extrapolated by several orders of magnitude (particle Lorentz factors, magnetic field values, electromagnetic spectra, etc.). Unfortunately, there is no consensus among research groups regarding the specific extrapolation method. Consequently, a universally accepted understanding of the physical origin of high-energy radiation from pulsars remains elusive.

Our proposal is to revisit fundamental principles and independently derive the ideal force-free magnetosphere using a novel numerical approach that treats the separatrix and equatorial discontinuities as genuine ideal current sheets.
In the present paper, we generalize in 3D the axisymmetric ideal solutions of \citet[][hereafter Paper~I]{2024MNRAS.528.3141D} and \citet[][hereafter Paper~II]{2024Univ...10..178C}. This is the third paper (Paper~III) of an ongoing investigation that aims to produce the reference ideal solution of the 3D pulsar magnetosphere.

\section{Domain decomposition}

Our goal is to obtain the steady-state structure of the inclined rotating pulsar magnetosphere with an independent new methodology. 
We address the problem in the mathematical frame that corotates with the star, keeping however the values of the magnetic and electric fields as measured in the laboratory (non-rotating) frame. We will work in both cartesian $(x,y,z)$ and spherical coordinates $(r,\theta,\phi)$ centered on the star and oriented along the axis of rotation. This is a mathematical system of coordinates corotating with the star, and there are no Lorentz transformations between that frame and the laboratory frame. The problem has been formulated as such by \citet{2009ApJ...692..140M} \citep[see also][for further analysis]{2012MNRAS.424..605P}, and the requirement of steady-state is equivalent to setting $\partial/\partial t=0$ in the corotating frame. Following the Muslimov and Harding formulation it is straightforward to show that, in steady state,
\begin{equation}
{\bf E} \equiv {\bf E}_p = -\frac{r\sin\theta}{R_{\rm lc}}\hat{\phi}\times {\bf B}\equiv -\frac{r\sin\theta}{R_{\rm lc}}\hat{\phi}\times {\bf B}_p\ .
\end{equation}
Here, the subscript $p$ denotes a poloidal field component (i.e. only the $r$ and $\theta$ components of the field). $R_{\rm lc}\equiv c/\Omega$ is the radius of the so-called light cylinder. It is interesting that, in steady-state, $E_\phi=0$ even in the general 3D problem. Furthermore, the force-free condition
\begin{equation}
\rho_e{\bf E}+{\bf J}\times {\bf B}=0
\label{ff1}
\end{equation}
may be rewritten in steady-state as
\begin{equation}
\nabla\times\left\{\left(1-\left[\frac{r\sin\theta}{R_{\rm lc}}\right]^2\right){\bf B}_{\rm p}+{\bf B}_\phi\right\}-\alpha{\bf B}=0\ ,
\label{ff2}
\end{equation}
where $\alpha$ is an electric current  function that is constant along individual field lines, namely
\begin{equation}
\nabla\alpha\cdot {\bf B}=0\ .
\label{alpha1}
\end{equation}
Eq.~(\ref{ff2}) was first obtained by \citet{1974ApJ...187..359E}  and \citet{1975MSRSL...8...79M} but has never been solved before in 3D. 

Solving in the co-rotating frame has an important advantage over time-dependent simulations in the non-rotating (laboratory) frame. In the latter, the final configuration is rotating, i.e. it is time-dependent. All its complex features (current sheets, Y-point, etc.) rotate in the simulation frame of reference, hence it is difficult to guarantee that a steady-state is truly reached. On the other hand, solving for the steady-state configuration in the rotating frame is a relaxation approach that more naturally reaches the final steady state. In that frame, current sheets develop at fixed positions of the numerical grid. 

The new methodology, first proposed in Paper~I to avoid the mathematical problems associated with the magnetospheric current sheet, is the domain decomposition into open and closed field lines (see schematic in figure~1 of that paper). In particular, we choose from the very beginning of our simulation which field lines will be open and which ones will be closed. This is equivalent to an ad hoc determination from the very beginning of the extent of the polar cap (see below). In general, the solution that we will obtain will contain a closed-line region that does not extend all the way to the light cylinder. \textcolor{black}{Obviously, the choice of the extent and shape of the polar cap is arbitrary, thus the solution is degenerate, as in the axisymmetric analyses of \citet{2005A&A...442..579C} and \citet{2006mn.368.1055T}.}

\textcolor{black}{In the early literature \citep[i.e. before the era of ab-initio global PIC simulations;][]{2015ApJ...801L..19P,2018ApJ...857...44K} it was expected that the single solution that nature will choose is the one in which the closed-line region has its maximum extent and everywhere touches the light cylinder. We have seen, however, in the PIC literature of the past 10 years that this is not the case and that the closed-line region ends some significant distance inside the light cylinder. \citet{contopoulos2024pulsar} offered a tentative explanation for this effect. The issue of the extent of the closed-line region is further complicated after Paper~II who concluded that, when the closed-line region reaches the light cylinder, the whole magnetosphere is contained inside the light cylinder, the open-line region disappears altogether, and the pulsar dies out completely. Several numerical simulations exist in the literature in which the closed-line region touches the light cylinder \citep[e.g.][etc.]{1999ApJ...511..351C,2006mn.368.1055T,2006ApJ...648L..51S}, thus we need to understand the origin of this discrepancy.}

\textcolor{black}{Once the size and shape of the polar cap are determined, the separatrix surface is also determined. Initially, however, its shape is unknown to us.} We then implement in 3D the procedure first proposed in Paper~I. We begin by separating the regions of open and closed field lines with an ad hoc chosen surface, namely
\begin{equation}
r_{\rm S}=r_{\rm S}(\theta,\phi)\ ,
\label{rS}
\end{equation}
that originates at the edges of the polar caps around the magnetic poles of the central star. \textcolor{black}{We will denote this surface the `interim separatrix'. As we will see below, we will iteratively adjust its shape above the polar cap till it reaches the final shape of the true separatrix that corresponds to the fixed size and shape of the polar cap that we have chosen.} 
A natural (but certainly not unique) initial choice might be a dipolar surface that ends on a circular polar cap centered around the magnetic axis. The reason we decided to separate the two regions is that, due to the electric current sheet that flows along the separatrix between the two regions, a strong contact discontinuity
(of practically infinitesimal thickness) in the magnetic and electric fields develops along the separatrix. \citet{2003ApJ...598..446U} integrated eq.~(\ref{ff1}) across current sheets and obtained that
\begin{equation}
(B^2-E^2)_{\rm below} = (B^2-E^2)_{\rm above}
\label{B2E2}
\end{equation}
 (see also Lyubarskii 1990). Treating such discontinuities inside any computational domain
is very problematic and all computational methods generate
spurious Gibbs oscillations around the separatrix \citep[see discussion in][]{2022ApJ...925..130C}. We propose here a new way to treat such discontinuities as follows:
\begin{enumerate}
\item Solve eqs.~(\ref{ff2}) and (\ref{alpha1}) independently in the two regions for an initial arbitrary choice of the \textcolor{black}{interim separatrix} between them. 
\item Implement boundary conditions for $B_r$ on the stellar surface (we do not impose boundary conditions on $B_\theta$ and $B_\phi$ since that would overdetermine the problem).
\item Implement the condition that the \textcolor{black}{interim separatrix} lies along magnetic field lines,  namely $\hat{\bf n}\cdot {\bf B}=0$ right above and below it, where
$\hat{\bf n}$ is a vector perpendicular to the \textcolor{black}{interim separatrix}.
\item Implement the requirement that $\nabla\cdot{\bf B}=0$.
\item Implement special boundary conditions at large radial distances (e.g. $B_r > 0$, $B_\theta/B\rightarrow 0$, $E_\theta/B_\phi\rightarrow 1$ etc. as $r\rightarrow \infty$) to help the convergence of the numerical method.
\item Adjust the shape of the \textcolor{black}{interim separatrix} so as to gradually remove the discontinuity of $B^2-E^2$ over its surface ($B^2-E^2$ will in general be discontinuous for an ad hoc choice of the \textcolor{black}{interim separatrix}; see Papers~I \& II for details). 
\item Repeat all of the above steps after each re-adjustment of the \textcolor{black}{interim separatrix} till pressure balance is satisfied at all points and the \textcolor{black}{interim separatrix} does not need to re-adjust its shape anymore. The \textcolor{black}{true separatrix} and the final steady-state solution will thus be obtained.
\end{enumerate}

We have implemented one more trick that greatly simplifies our problem, namely a novel
numerical treatment of the equatorial current sheet that originates at the tip of the closed line region, as first proposed in Paper~II. The reason there exists an equatorial current sheet is that magnetic field lines leave one pole of the star, open up to infinity, and return from infinity to the other pole of the star. In doing so, they also carry the same amount of poloidal electric current in each hemisphere from each pole of the star to infinity. These two electric currents return to the star through the equatorial current sheet. In other words, the only reason an equatorial current sheet develops is simply to satisfy closure of the global poloidal electric current circuit \citep{1999ApJ...511..351C}. It is thus obvious that, if we artificially (numerically) invert the direction of the field lines that leave the star from the southern pole, the electric current direction in the southern hemisphere will be inverted, and therefore there will be no need to close the global poloidal electric current circuit through an equatorial current sheet. This configuration is clearly artificial (it is equivalent to a magnetic monopole), but it is mathematically and dynamically equivalent to the configuration that we are investigating in the open line region, only without the mathematical discontinuity of the equatorial current
sheet! We are able to implement this trick because we are treating the open line region independently from the closed line region. This is the same trick assumed by Bogovalov (1999) when he obtained the solution for the tilted split monopole. We here generalize his approach and show that it is also valid (and very helpful) in the numerical treatment of the open line region in the more general dipole magnetosphere. We have thus found a way to make the equatorial current sheet discontinuity ‘numerically disappear’. The undulating surface separating open magnetic field lines that originate in the north stellar hemisphere from those that originate in the south is where the equatorial current sheet truly lies when the magnetic field direction in the southern stellar hemisphere is reversed back to its true value.


\begin{figure}
		\centering
		\includegraphics[width=1.\columnwidth]{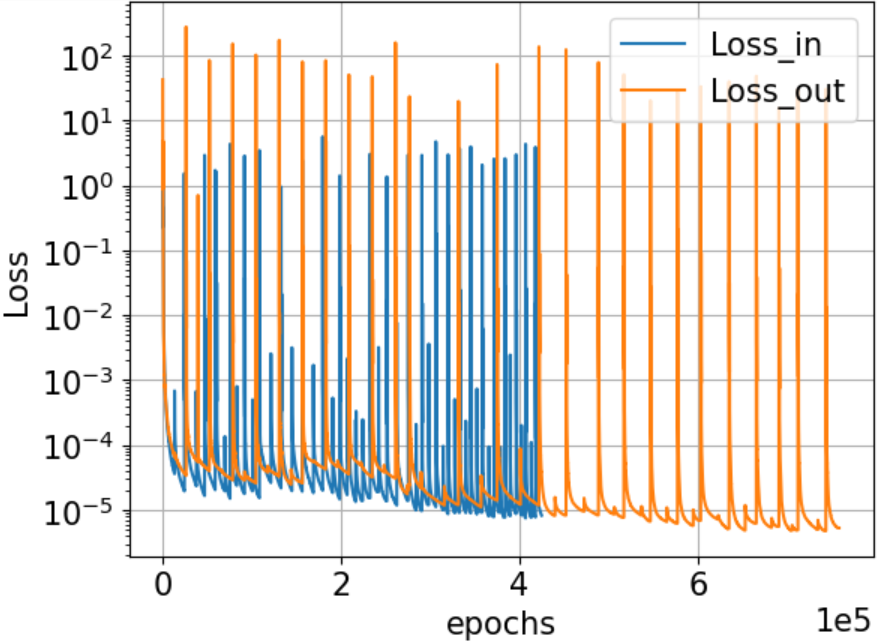}
		\caption{Evolution of the total NN losses for the solution with $\lambda=20^\circ$  and $\theta_{\rm pc}=36^\circ$ in the closed-line (blue) and open-line (orange) regions with respect to the number of training epochs (the difference in the number of epochs has to do with the different number of steps the training algorithm takes during its second-order optimization). We started the training with a first order optimizer and switched to a second order one after $10^4$ training epochs. Beyond that, we adjusted the shape of the \textcolor{black}{interim separatrix} every $3\times 10^4$ epochs \textcolor{black}{(blue and orange high spikes)}.
The achieved total training loss between $10^{-5}$ and  $10^{-6}$ is deemed satisfactory.}
		\label{fig:losses}
	\end{figure}

\begin{figure}
		\centering
\includegraphics[width=1.1\columnwidth]{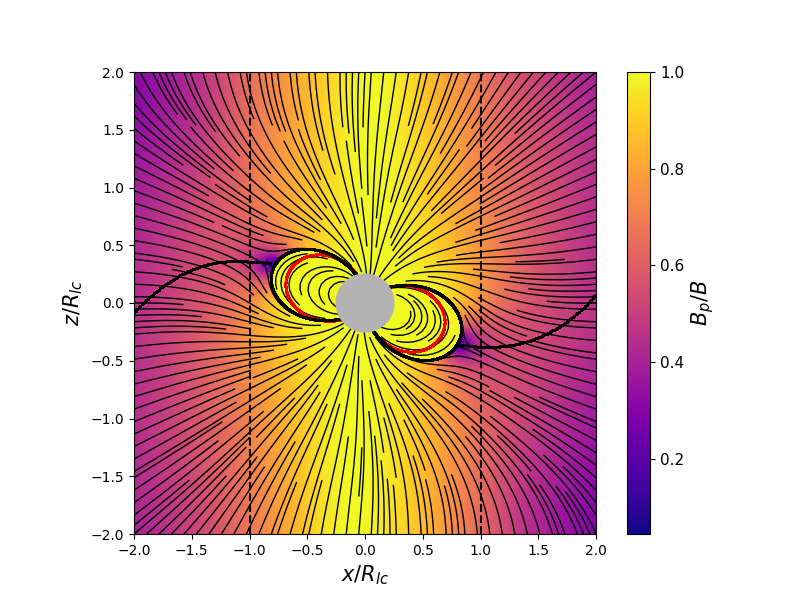}
\caption{Cross section of steady-state solution representing an inclined rotator with $\lambda=20^\circ$ and $\theta_{\rm pc}=36^\circ$. In this simulation $r_*=0.25 R_{\rm lc}$. Rotation axis along $z$. The inclined magnetic axis lies along the corotating $xz$ plane shown. {This particular cross section represents phase $0$ (or $\pi$) of the corresponding rotating time-dependent solution.}
{Closed thick black lines: separatrix between open and closed field lines. Open thick black lines: separatrix between open field lines that originate from the north and south magnetic poles. This is where the equatorial current sheet lies}. Red lines: initial dipolar shapes of the \textcolor{black}{interim separatrices} before readjustment. For this particular choice of the polar cap, the dipole is significantly stretched outwards closer to the light cylinder (represented by the two dashed lines  at $x/R_{\rm lc}=\pm 1$). 
The equatorial current sheet lies where lines from the north and south polar cap rim meet. Color scale: ratio $B_p/B$. This represents the development of the azimuthal magnetic field $B_\phi$ accross the magnetosphere. Notice that at the magnetospheric Y-point where the equatorial current sheet connects to the separatrix current sheet, $B_p=0$ and $B_\phi\neq 0$ as expected \citep{2003ApJ...598..446U}. \textcolor{black}{Notice also that  in this solution the Y-point is still significantly inside the light cylinder.}}
		\label{fig:magnetic_solution0}
	\end{figure}
\begin{figure}
		\centering
\includegraphics[width=1.1\columnwidth]{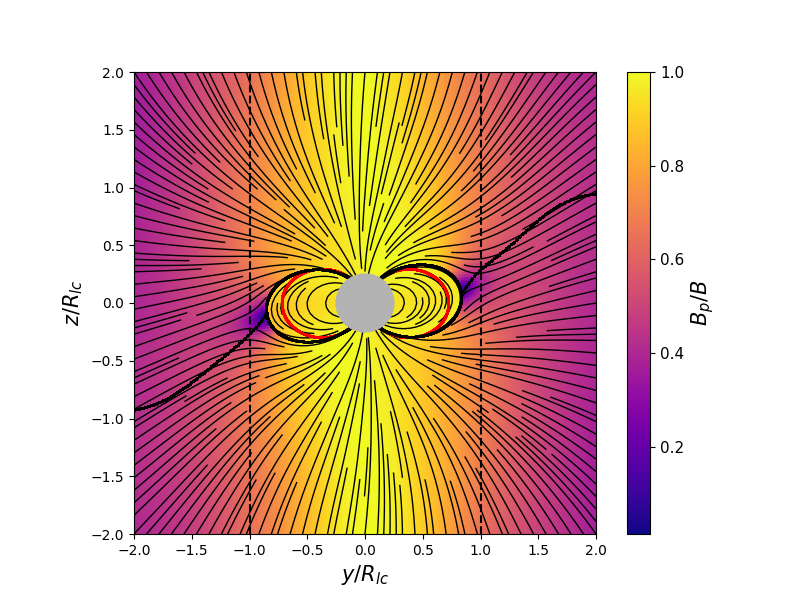}
\caption{Same as Fig.~\ref{fig:magnetic_solution0} but along the corotating $yz$ plane. {This particular cross section represents phase $\pi/2$ (or $3\pi/2$) of the corresponding rotating time-dependent solution.}}
		\label{fig:magnetic_solution02}
	\end{figure}

\begin{figure}
		\centering
\vspace{0.2cm}
\includegraphics[width=1\columnwidth]{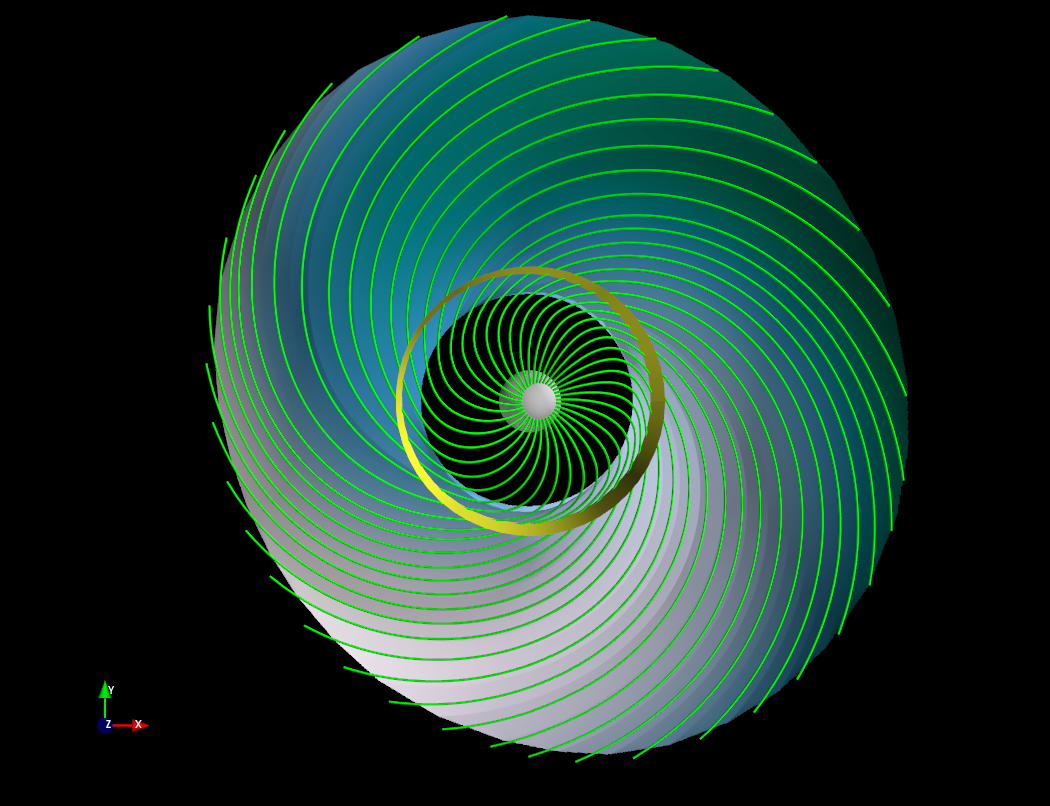}
\caption{The steady-state solution of Fig.~\ref{fig:magnetic_solution0} in 3D as seen from above. Yellow cylinder: light cylinder. {Pale undulating surface: equatorial current sheet}. We plot only open magnetic field lines just inside the rim of the northern polar cap. \textcolor{black}{The Y-point in this solution is significantly inside the light cylinder.}
We see the clear azimuthal break of open field lines expected very close to the Y-point where $B_p\rightarrow 0$ and $B_\phi \neq 0$. 
}
		\label{fig:magnetic_solution1}
	\end{figure}
\begin{figure*}
		\centering
\includegraphics[width=2\columnwidth]{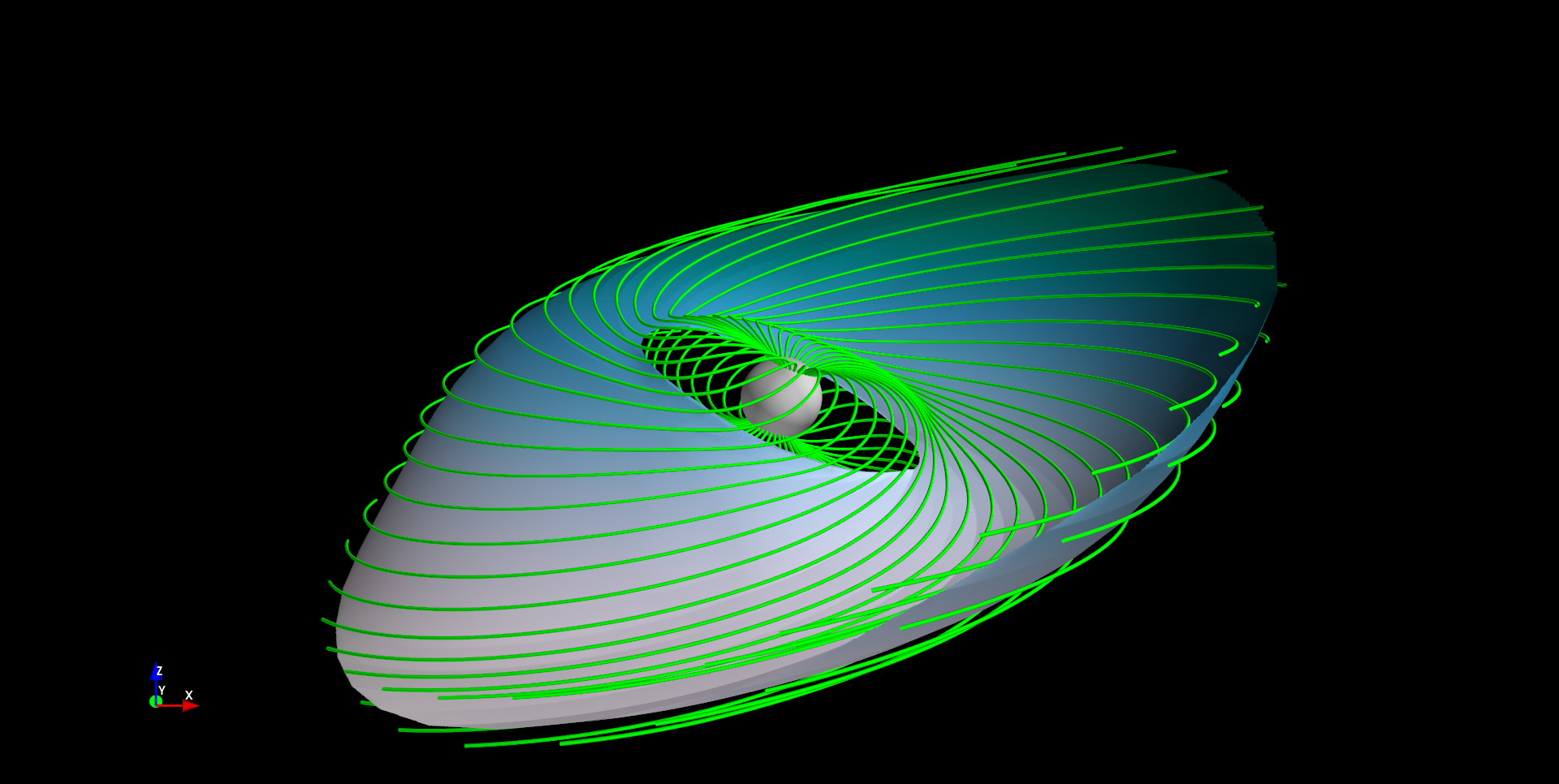}
\caption{The steady-state solution of Fig.~\ref{fig:magnetic_solution1} as seen from the side. {We plot open magnetic field lines just inside the rims of both polar caps}.The undulating shape of the equatorial current sheet between them is clearly seen. 
{Notice that it does not develop the physical instabilities seen in all previous numerical solutions because this physics is missing in our work (we obtain the solution without a current sheet and then reverse the direction of the field in the southern magnetic hemisphere).}
}
		\label{fig:magnetic_solution11}
	\end{figure*}


\begin{figure}
		\centering
\includegraphics[width=1\columnwidth]{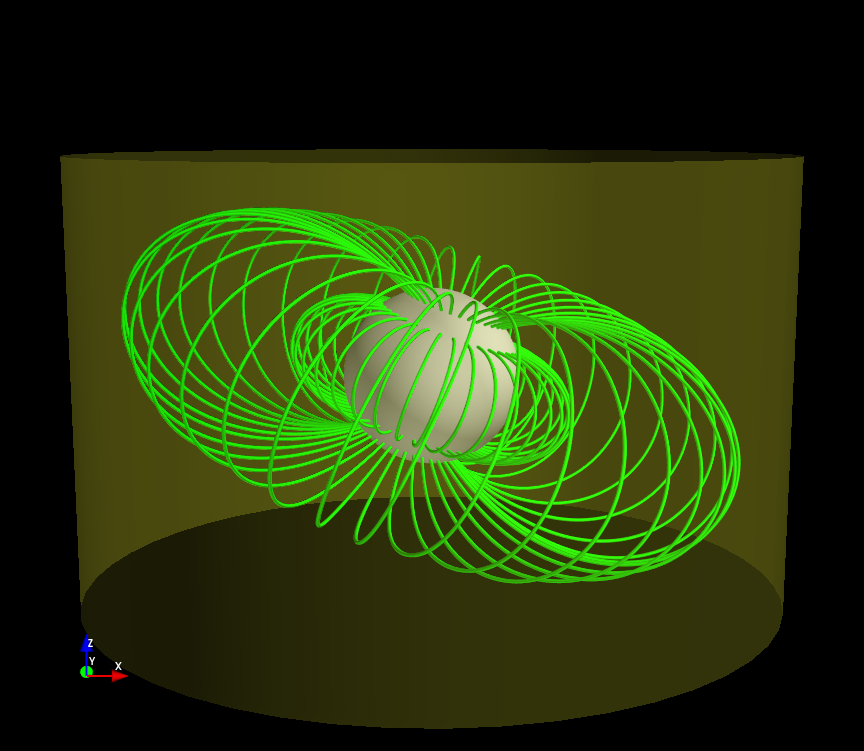}
\caption{Closeup of Fig.~\ref{fig:magnetic_solution11} near the stellar surface in the closed line region. 
In this solution we required that $\alpha=0$ in that region. Nevertheless, we observe that field lines develop a clear azimuthal twist with respect to the magnetic axis.}
		\label{fig:magnetic_solution2}
	\end{figure}

\begin{figure}
		\centering
\vspace{0.25cm}
\includegraphics[width=1\columnwidth]{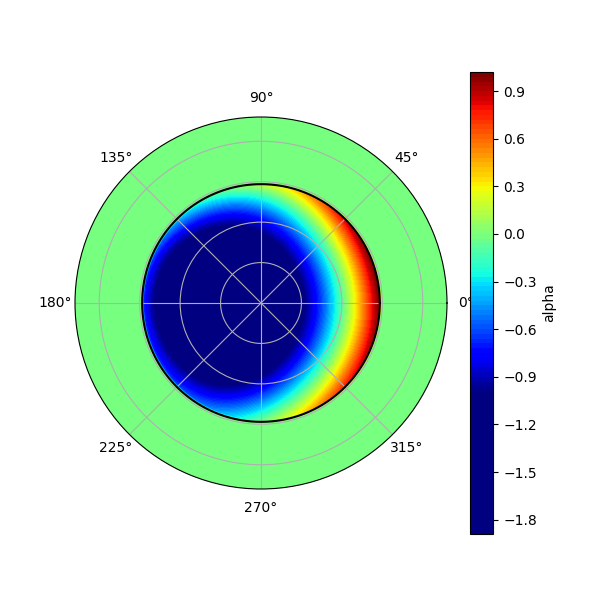}
\caption{Distribution of current parameter $\alpha$ along the stellar surface {as seen from above the 
polar cap} for the solution of Fig.~\ref{fig:magnetic_solution0}. $\alpha=0$ in the green closed-line region outside the polar caps. {$\phi=0$ along the corotating $xz$ plane of Fig.~\ref{fig:magnetic_solution0} in the direction of inclination of the magnetic axis}. Blue region: main magnetospheric current. Yellow-red region: part of the return current (the main part of the return current lies along the separatrix-thick black dotted line).
}
		\label{polarcap}
	\end{figure}


\begin{figure}
		\centering
\includegraphics[width=1\columnwidth]{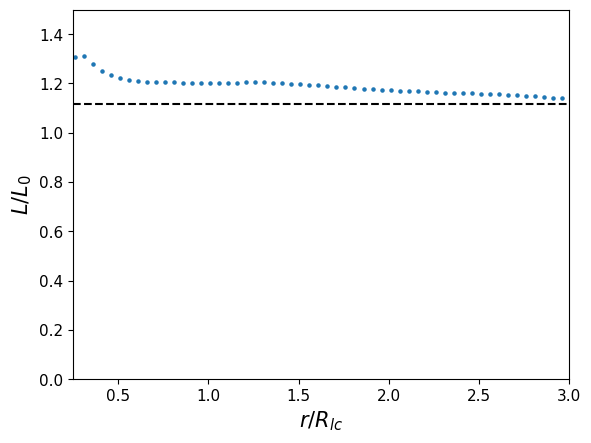}
\caption{Evolution with distance of the total Poynting flux $L$ calculated over spheres of radius $r$ centered over the central star, normalized with respect to the aligned rotator's canonical luminosity value $L_0\equiv B_*^2 r_*^6 c/(4R_{\rm lc}^4)$. The value of $L$ {is close to} previous estimates in the literature (dashed line according to \citet{2006ApJ...648L..51S} for the  pulsar inclination angle considered). Energy is (almost) conserved beyond the Y-point and the light cylinder. 
}
		\label{fig:magnetic_solution3}
	\end{figure}

Requirement (1) corresponds to three equations for the three components of the force-balance equation and one more equation for eq.~(\ref{alpha1}), while  requirements (2), (3), (4) \& (5) add four more equations. All equations are written in cartesian coordinates in order to avoid the singularity of spherical coordinates around the axis $\theta=0$ and the requirement for periodicity between $\phi=0$ and $\phi=2\pi$. Notice that no central symmetry is required by the solution.
All these requirements were satisfied with Machine Learning techniques (hereafter ML), in particular standard Fully Connected Neural Networks (hereafter NNs). NNs have an advantage in that they are meshless and thus can be trained with ease over complex deformable domains, and a disadvantage in that they occasionally fail to converge to the absolute optimization minimum. Physics Informed NNs (hereafter PINNs) are ordinary NNs trained over loss functions that are related to a physical problem. In particular, we trained our PINNs with a loss function that consists of the 8 individual requirements discussed above that must all be minimized to zero. Notice that each part consist of the square of the expression that we need to minimize, averaged over all its corresponding training points (i.e. in the exterior and interior pulsar magnetosphere, along the separatrix between the two regions, along the stellar surface, and at the outer boundary of our simulation at large $r$). NNs are proven to be very capable optimizers and can satisfy to sufficient precision all of the above constraints. 

\section{Results}

We experimented with various NN configurations, number layers and nodes per layer, various optimization procedures, and various activation functions. This tuning of hyperparameters is a tedious manual process that requires significant effort before the first meaningful results are obtained. In the solution presented below we implemented two Fully Connected NNs with 3 entires ($x, y, z$), 5 internal layers consisting of 128 nodes each, and 4 exits ($B_x, B_y, B_z, \alpha$), first order Adam optimizers initially and second order Broyden-Fletcher-Goldfarb-Shanno (BFGS) optimizers later on, tanh activation functions, and roughly 270,000 training points distributed randomly over the various magnetospheric computational domains and their boundaries. Notice that, in order to avoid overfitting, the training points are periodically updated. 
The reason we implemented a cartesian system of coordinates was to avoid the singularities of the spherical system of coordinates around the axis $\theta=0$ and the requirement for periodicity between $\phi=0$ and $\phi=2\pi$. We validated our procedure by training our NNs with the analytic vacuum dipole solution of $\nabla\times {\bf B}=0$ (no rotation, no light cylinder, no electric currents). The comparison with the analytic  dipolar solution was fair (see also Paper~I). This gave us confidence to address the full 3D pulsar magnetosphere problem with PINNs. 
{In Fig.~\ref{fig:losses} we show the evolution of the total training losses for our main NNs. The shape of the \textcolor{black}{interim separatrix} was re-adjusted every 30,000 training epochs. We performed 25 readjustments (high spikes in Fig.~\ref{fig:losses}) in order to reach a solution with individual requirements satisfied to a precision of a few times  $10^{-6}$ in both NNs. The solution relaxes to a particular separatrix shape that satisfies pressure balance at all points accross it. The training of our two NNs required about 24 hours of computational time on a standard off-the-shelf computer with a GPU.}
We thus obtained a dissipationless ideal steady-state solution for a star with radius $r_*=0.25 R_{\rm lc}$, a uniform polar cap with constant angular opening $\theta_{\rm pc}=36^\circ (\pi/5)$ around the magnetic axis, and an inclination angle of $\lambda=20^\circ (\pi/9)$. As we will see next, our first results are promising and emphasize the potential of our method. 

{In Figs.~\ref{fig:magnetic_solution0} and \ref{fig:magnetic_solution02} we show two cross-sections of the 3D solutions along the corotating $xz$ plane that contains both the rotation and magnetic axes, and along the corotating $yz$ plane perpendicular to it. We observe that the initially dipolar separatrices are significantly stretched  closer to the light cylinder. 
We also observe a clear nulling of the poloidal field $B_p$ at the Y-point where $B_\phi$ is non-zero due to the return current that flows back to the star along the equatorial current sheet. This result was predicted by \citet{2003ApJ...598..446U} but has never been seen before so clearly, mainly because no previous ideal force-free solution obtained zero-thickness equatorial and separatrix current sheets.  \citet{2003ApJ...598..446U} also predicted that, because $B_p=0$ and $B_\phi\neq 0$ right outside the Y-point, while $B_\phi\approx 0$ right inside, in order to achieve pressure balance between the inside and the outside at that point, $B_p$ must be non-zero just inside the Y-point. Therefore,  technically, the shape of the Y-point must be a T as is clearly observed in Figs.~\ref{fig:magnetic_solution0} and \ref{fig:magnetic_solution02}. This result too has never been seen before in all previous numerical solutions, starting from \citet{1999ApJ...511..351C}, in which the separatrix is always shown with a pointed shape where the equatorial current sheet originates.

In Figs.~\ref{fig:magnetic_solution1}, \ref{fig:magnetic_solution11} we show the 3D distribution of open magnetic field lines right above {and below} the magnetospheric equatorial and separatrix current sheet.
Notice that the value of $\alpha$ is numerically adjusted along each open field line so that the latter smoothly crosses the light cylinder. It is interesting that while \citet{1999ApJ...511..351C} developed a complex numerical method to perform this adjustment, NNs are able to accomplish this automatically. 
In Fig.~\ref{fig:magnetic_solution11} we clearly observe the undulating shape of the equatorial current sheet beyond the closed line region. {This is determined as the boundary between the regions of open field lines that originate from the north and south magnetic poles. Notice that the undulating black lines in Figs.~\ref{fig:magnetic_solution0} and \ref{fig:magnetic_solution02} were obtained from the intersection of the 3D undulating current sheet of Figs.~\ref{fig:magnetic_solution1} and \ref{fig:magnetic_solution11} with the $xz$ and $yz$ planes respectively.}
}

In Fig.~\ref{fig:magnetic_solution2} we observe that closed field lines develop a higher azimuthal twist with respect to the magnetic axis the closer they are to the separatrix. Nevertheless, in the closed line region we required that $\alpha=0$. 
Apparently, in order for electric currents not to flow along them, closed field lines must connect footpoints that have the same electric potential. In the case of an inclined rotator, such footpoints do not lie on the same stellar meridional as in the case of an axisymmetric rotator. If indeed true, this is an interesting result that needs to be further investigated. The footpoints of closed magnetic field lines are not connected between themselves in a non-rotating inclined dipole. This makes us wonder how do closed magnetic field lines that were initially dipolar evolve in a time-dependent ideal simulation that involves the rotation of an initially static inclined dipole \citep[as e.g. in][and thereafter]{2006ApJ...648L..51S}. An ideal time-dependent simulation should hold the same footpoints for ever. It is not clear to us how do these footpoints evolve in all previous time-dependent ideal simulations. 
The distribution of the electric current parameter $\alpha$ is shown in Fig.~\ref{polarcap} where we see the neutron star surface from above the rotation axis. The main magnetospheric current flows along the blue region where $\alpha<0$ for a pulsar with inclination $\lambda< 90^\circ$ as in the present solution. The return current flows mainly along the separatrix (black dotted line), and partly in the yellow-red region inside the polar cap  \citep[compare with figure~4 of][]{2010ApJ...715.1282B}.

In Fig.~\ref{fig:magnetic_solution3} we calculate the integral of the 
Poynting vector ${\bf E}\times {\bf B}/(4\pi)$
over spheres of radius $r$ centered around the central star. We see that, because of the absence of numerical current sheets, the total Poynting flux is (almost) conserved in the open line region of our calculation, {and we observe a slight numerical dissipation of around $10\%$ up to about $3R_{\rm lc}$}. All previous numerical solutions show a clear drop of the integrated Poynting flux beyond the light cylinder due to numerical dissipation at the equatorial current sheet.
The obtained value of the total electromagnetic luminosity $L$ is comparable to the estimate of \citet{2006ApJ...648L..51S}, namely 
\begin{equation}
L=1.2\pm 0.05 \times L_0 (1+\sin^2\lambda)\ ,
\label{L}
\end{equation}
{from $r\approx 0.4 R_{\rm lc}$ to about $3R_{\rm lc}$}. Here $L_0=B_*^2 r_*^6 c/(4R_{\rm lc}^4)$ is the canonical aligned rotator's luminosity value, and $\lambda=20^\circ$ is the pulsar inclination angle. 
We will compute a complete sequence of pulsar inclination angles, polar cap openings, and polar cap shapes in Paper~IV. We will then learn whether the result of Paper~II is reproduced in 3D, namely whether magnetospheric solutions exist even for arbitrarily small polar caps with luminosities $L$ much smaller than the canonical value of eq.~(\ref{L}).


\section{Summary and conclusions}

{We have developed a novel robust numerical method that allows us to calculate steady-state force-free ideal 3D pulsar magnetospheres for various pulsar inclinations and polar cap shapes}. Our method {\it does not} involve a time-dependent simulation that eventually relaxes to a steady-state, as in all previous attempts to obtain the steady-state pulsar magnetosphere. Instead, we solve directly the generalization of the axisymmetric pulsar equation \citep{1973ApJ...182..951S} in 3D \citep{1974ApJ...187..359E,1975MSRSL...8...79M}. Notice that the method is not limited to centrally located dipole fields and can accomodate \textcolor{black}{more complex} stellar magnetic field configurations. 

\textcolor{black}{For example, one may determine a particular magnetic field configuration with hotspots that reproduce the NICER X-ray observations, assume that these hotspots correspond to the origin of open magnetic field lines (i.e. assume that these are the polar caps), and calculate the 3D ideal magnetosphere that corresponds to these particular polar caps using our methodology. 
The initial field configuration that would reproduce these polar cap shapes will require inclusion of higher multipoles. Very small polar caps can be produced by an aligned and displaced dipole and quadrupole \citep[e.g.][]{Gralla2017}, and NICER hot spots require more complicated field structures with off-set dipole and quadrupole configurations \citep{Chen2020,Kalapotharakos2021}. This will complicate our choice of the initial ad hoc shape of the interim separatrix, but as long as the polar caps remain fixed, the iteratively adjustable shape of the interim separatrix will remain topologically the same, thus we believe that our procedure is directly applicable and will indeed eventually relax to the true shape of the separatrix. 
}

Our methodology involves the decomposition of the pulsar magnetosphere into the regions of open and closed field lines. This allows us to treat the separatrix current sheet  as a perfect mathematical discontinuity between the two regions. We also `numerically remove' the equatorial current sheet by reversing the field direction in the southern hemisphere, and reverting it to its true direction after the 3D solution is obtained. {We do acknowledge that although this approach helps in dealing with magnetospheric current sheets numerically, it completely ignores the physics of reconnection, electromagnetic energy dissipation, plasma instabilities and particle acceleration along them. In that sense, our solutions may be considered only as ideal solutions that must be augmented with the microphysics of the current sheets.}
Our present method can be directly applied to more general ideal MHD problems that contain current sheets, as for example in active regions of the solar corona.
In this paper \textcolor{black}{we present our first solution for one particular size and shape (circular) of the polar cap of open magnetic field lines}, and for one particular pulsar inclination angle with respect to the axis of rotation. The solution is obtained in a coordinate system that corotates with the star. {Our results elucidate the potential of our method to generate new solutions of the ideal pulsar magnetosphere}. 

We would like to emphasize that the general steady-state ideal force-free solution is degenerate in the same sense that the axisymmetric solutions of  \citet{2005A&A...442..579C} and \citet{2006mn.368.1055T} are degenerate. In the axisymmetric case, the polar cap is circular, but its radial extent is arbitrary and depends on how far the closed-line region extends from the light cylinder (the larger the polar cap, the shorter the closed-line region). Obviously, the solution where the closed-line region touches upon the light cylinder is unique, and consequently, the corresponding extent of the polar cap is also uniquely defined. In our present 3D case, not only is the size of the polar cap arbitrary, but also its shape and position. The solution that we obtained in this work corresponds to a circular polar cap centered around the magnetic axis. 
\textcolor{black}{This choice causes the closed field line boundary to lie at different distances with respect to the light cylinder as a function of azimuth and therefore does not capture correctly the field line sweep back at the polar cap. With our chosen inclination of $20^\circ$, this does not make much difference in the solution, but for larger inclination angles, it will make a big difference.
}
We are certainly {\it not} claiming that this is the only ideal steady-state force-free solution of the pulsar magnetosphere. The general solution is clearly degenerate, and what we are presenting is just one valid solution. 
If we insist that the closed-line region everywhere touches upon the light cylinder, that solution will be unique and it will correspond to one unique polar cap shape that will not be circular nor centered around the magnetic axis. In fact, it has been shown in multiple studies that the polar caps on to which previous time-dependent numerical simulations converge are very different from those of a static (or even a retarded) vacuum dipole and depend on the magnetic inclination angle \citep[e.g.][]{Dyks_Harding2004,Timokhin_Arons2013}.

In this paper, we have trained a set of two NNs that instantaneously yields the three components $B_x,B_y,B_z$ of the magnetic field and their spatial derivatives at any given point $(x,y,z)$ of the magnetosphere for one particular pulsar inclination, and one particular polar cap size and shape. This is extremely helpful if one wants to draw magnetospheric features like magnetic and electric field lines, electric charge and electric current distributions, Poynting flux, current sheets, etc. 
We will present a detailed investigation of the pulsar magnetosphere for various pulsar inclinations and for various polar cap sizes and shapes in an extended forthcoming publication, and we will produce corresponding trained NNs (Paper~IV).
\textcolor{black}{We also plan to investigate the unique solution where the closed-line region everywhere touches upon the light cylinder. Determining this unique solution will require adjusting both the polar caps as well as the separatrix at every iteration step. 
}
We look forward to learn whether magnetospheric solutions exist even for arbitrarily small polar caps with arbitrarily small luminosities as seen in 2D (Paper~II).
We would like to emphasize that our methodology allows us to obtain new solutions of the ideal 3D pulsar magnetosphere by determining a priori {various shapes and sizes} of the polar cap.
Such solutions were never obtained before because all numerical solutions obtained in the literature always relax to one particular magnetospheric configuration, i.e. to one particular polar cap shape and size determined by the simulation, without any possibility for control by the programmer. This opens up a whole spectrum of solutions between which the real pulsar magnetosphere may choose to transition during its evolution. More work is needed along these lines.


\section*{Acknowledgements}

This work was supported by computational time granted from the National Infrastructure for Research and Technology S.A. (GRNET) in the National HPC facility - ARIS - under project ID pr016005-gpu. ID is supported by the Hellenic Foundation for Research and Innovation (HFRI) under the 4th Call for HFRI PhD Fellowships (Fellowship Number: 9239). \textcolor{black}{EC is supported by the Sectoral Development Program (${\rm O}\Pi\Sigma$ 5223471) of the Ministry of Education, Religious Affairs and Sports of Greece, through the National Development Program (NDP) 2021-25.}

\section*{Data Availability}
The data underlying this article will be shared on reasonable re-
quest to the corresponding author.

\bibliographystyle{aa}
\bibliography{Literature.bib} 

\section*{Appendix}

We offer here a tentative explanation why, in all PIC simulations of the past decade, the separatrix current sheet turns out to be so much thicker than the equatorial current sheet. 

Both current sheet are charged due to the discontinuity of the vertical electric field component across them (which is itself due to the discontinuity of the poloidal magnetic field, that is itself due to the presence of poloidal electric currents only in the open-line region). We can thus also call them `charge sheets'. The question is what is the source of the electric charges needed to support both the charge and the electric current of the sheet. Regarding the equatorial current sheet, it is well accepted that relativistic reconnection takes place all along it, therefore, magnetospheric field lines enter it over an extended region beyond the Y-point. Each such field line carries electron-positron pairs with it which are thus introduced in the equatorial current sheet. Therefore, the supply of charges in the equator is ample. As a result, the equatorial current sheet can be really thin as is indeed seen in all previous PIC simulations.

This is not the case, however, with the separatrix surface. Reconnection along it is minor because this is not a Harris-type current sheet as the equatorial one (it contains a strong guide field), and there are no field lines entering it along the way from the stellar surface to its tip at the Y-point. One might think that the high density of magnetospherically supplied pairs that allowed the equatorial current sheet to appear so thin in the numerical PIC simulations will also supply the necessary pairs in the separatrix. Unfortunately these are not enough since, as we have shown in \citet{2020MNRAS.491.5579C}, an extra outward flow of positrons/protons is required through the separatrix to support the separatrix electric current and electric charge\footnote{Without them, the negative surface charge density at the footpoint of the separatrix on the polar cap due to the electrons inflowing from the Y-point would be orders of magnitude higher than the expected electric  field discontinuity accross the separatrix at that position.}. This flow of positive charges must originate on the stellar polar cap, where their charge density must accordingly be immense (a positronic current sheet corresponds also to a positronic charge sheet of infinite density-mathematically). In all previous PIC simulations, however, only a finite pair density on the order of the Goldreich-Julian value is introduced everywhere on the stellar surface. This is insufficient to support a current/charge sheet, therefore, the separatrix electric current is distributed over a wide region of poloidal field lines on the stellar surface. This effect eliminates the concept of a `separatrix surface' and leads to the strange features observed in all previous PIC simulations, namely a thick separatrix and a region of strong dissipation just inside the light cylinder.
We suspect that, if one introduces a much higher multiplicity of pairs on the stellar surface (e.g. 100 times the Goldreich-Julian value), the separatrix return current region will be correspondingly thinner \citep[e.g. 100 times; see also][]{2020MNRAS.491.5579C}. This would be a very interesting numerical experiment to perform that would justify our present approach of dividing the magnetosphere in the two regions of open and closed field lines separated by a separatrix surface (not the thick separatrix region of PIC simulations).

\end{document}